\documentclass[aps,prd,preprint,superscriptaddress]{revtex4}

\usepackage{graphicx}
\usepackage{dcolumn}
\usepackage{bm}
\usepackage{amsmath}
\usepackage{amssymb}
\usepackage{latexsym}
\usepackage{epsfig}
\usepackage{amsbsy}
\usepackage{array}
\usepackage{amssymb}
\usepackage{setspace}
\usepackage{bm}
\usepackage{graphicx}
\usepackage{subfigure}

\def\sint{\ifmmode{- \!\!\!\!\!\! \int}
    \else{\hbox{$- \!\!\!\! \int \ $}}\fi}

\begin{document}


\title{The nonleptonic charmless decays of $B_c$ meson }

\author{Wan-Li~Ju,\footnote{wl\_ju\_hit@163.com}~Tianhong Wang,~Yue~Jiang,~Han~Yuan, Guo-Li Wang\footnote{gl\_wang@hit.edu.cn}}
\affiliation{Department of Physics, Harbin Institute of Technology,
Harbin, 150001, China}


\begin{abstract}
In this paper, with the framework of (p)NRQCD and SCET, the processes $B_c\to M_1 M_2$ are investigated. Here $M_{1(2)}$ denotes the light charmless meson, such as $\pi$, $\rho$, $K$ or $K^*$. Based on the SCET power counting rules, the leading transition amplitudes  are picked out, which include $A_{wA}^2$, $A_{wB}^2$, $A_{wC}^2$, $A_{wD}^2$ and $A_{c}^0$.
From SCET,  their factorization formulae are proven. Based on the obtained factorization formulae, in particular,
the numerical calculation on $A_{wB}^2$ is performed.
%
%
%
%
%
%
%
%

\end{abstract}


\maketitle

\section{Introduction}\label{INtroductions}
Within the Standard Model, the $B_c$ meson is the only pseudo-scalar meson formed by two different heavy flavor quarks. Due to its mass being under the $B$$D$ threshold and the explicit flavors, $B_c$ meson decays weakly but behaves stably via the strong and electromagnetic interactions. Its weak decay modes are expected to be rich, because  the $B_c$ meson contains two heavy quarks. Either can decay independent, or both of them  annihilate to a virtual $W$ boson.

In the recent decades, the decays of $B_c$ meson have been widely studied. In this work, we lay stress on the two-body  charmless processes $B_c\to M_1 M_2$.
These charmless decays have particular features. First, they are not influenced by the penguin diagrams, which are expected to be sensitive to the new physics. Thus, they provide pure laboratories to examine the QCD effective methods.  Second, they receive the contributions only from the annihilation amplitudes, which offer an ideal opportunity to study the annihilation effects singly.


In the  paper~\cite{DescotesGenon:2009ja}, the nonleptonic charmless $B_c\to M_1 M_2$ decays have been calculated within the ``QCD Factorization'' approach (QCDF), while in Refs.~\cite{Liu:2009qa,Yang:2010ba}, these processes are calculated
in the  ``perturbatve QCD" (pQCD) schememethod.
However, in this work,
a sequence of effective field theories are employed to analysis the $B_c\to M_1 M_2$ transitions.
Considering that the initial meson of the $B_c\to M_1 M_2$ transitions is $B_c$, which include two heavy quarks, we use the non-relativistic effective theory of QCD (NRQCD) \cite{Bodwin:1994jh,Luke:1999kz} to deal with them.
Due to the relationship $M_{B_c}\gg M_{M_1}\sim M_{M_2}$, which makes that  the final mesons are relativistically boosted and back-to-back move, we use soft collinear effective theory (SCET) \cite{Bauer:2000ew,Bauer:2000yr,Bauer:2001ct,Bauer:2001yt,Bauer:2003mga,Bauer:2002aj} to describe these degrees of freedom (DOF). Under the SCET, it is convenient to explore the factorizations properties of the transition amplitudes.

%
%

This paper is organized as follows. In Sec. \ref{theoretical}, we introduce the theoretical details. We classify the transition amplitudes and focus on leading contributions. Within the framework of SCET, we prove the factorization formulae. Within Sec.~\ref{phno}, according to the  obtained factorization formulae, we calculate $A_{wB}^2$ and present the numerical results.

%
%
%

\section{Theoretical Details}\label{theoretical}
In this section, we present the theoretical details. First of all, the general frameworks are shown and the transition amplitudes are classified into categories, $A_w$ and $A_c$. Next, we pick out the leading contributions of $A_w$ and $A_c$, respectively and prove the according factorization formulae.
\subsection{Frameworks and Power Counting Rules}\label{Sec:powercounting}

As to the $B_c\to M_1M_2$ processes, there are three typical scales, $m_b$, $\sqrt{m_b\Lambda_{H}}$ and $\Lambda_{H}$. $\Lambda_{H}$ is the typical hadronic scale. Conventionally, $\Lambda_{H}\sim 500~\text{MeV}$ \cite{Bauer:2003mga}.

In order to describe the DOFs at scales $\sim m_b$, we use the full QCD and low-energy  effective Hamiltonian \cite{Buchalla:1995vs}, which is
\begin{equation}
H_W=\frac{2G_F}{\sqrt{2}}\sum_{q=d,s}V_{cb}V^*_{uq}(C_{1}\bar{c}_L\gamma^{\mu}b_L\bar{q}_L\gamma_{\mu}u_L+C_2\bar{c}_{L\beta}\gamma^{\mu}b_{L\alpha}\bar{q}_{L\alpha}\gamma_{\mu}u_{L\beta})+h.c.
\label{weakeffectiveHamiltonian}\end{equation}
In Eq.~\eqref{weakeffectiveHamiltonian}, $G_F$ denotes the Fermi coupling constant and  $V_{q_1q_2}$s stand for the CKM matrix elements.
$C_{1(2)}$ is the Wilson coefficients. $\mu$ represents the Lorentz index, while $\alpha(\beta)$ is the color index.


For investigating the  $\sqrt{m_b\Lambda_{H}}$ fluctuations, we need to
integrate out the hard modes $\sim m_b$, obtaining several  transition currents $J^{\text{I}}$ and the intermediate effective theory  $\text{SECT}_{\text{I}}+\text{NRQCD}$.
Within $\text{SECT}_{\text{I}}$~\cite{Bauer:2000ew,Bauer:2000yr,Bauer:2001ct,Bauer:2001yt}, there are three kinds of DOFs \cite{Bauer:2000yr}: 1) the $n$-collinear quarks $\xi^{\text{I}}_n$ and gluons $A^{\text{I}}_n$ with the momentum scaling $p_c=(n\cdot p_c,\bar{n}\cdot p_c, p_{c\bot})\sim m_b(\lambda^2,1,\lambda)$; 2) the $\bar{n}$-collinear quarks $\xi^{\text{I}}_{\bar{n}}$ and gluons $A^{\text{I}}_{\bar{n}}$ with the momenta $p_{\bar{c}}\sim m_b(1,\lambda^2,\lambda)$; 3)  the ultra-soft quarks $\xi^{\text{I}}_{us}$ and gluons $A^{\text{I}}_{us}$ with $p_{us}\sim m_b(\lambda^2,\lambda^2,\lambda^2)$.
$\lambda=\sqrt{\Lambda_{H}/m_b}$ is the expansion parameter. The power counting rules for these $\text{SECT}_{\text{I}}$ fields \cite{Bauer:2000yr} are summarized in Table.~\ref{powercountingSCET}.

\begin{table}[!htbp]
\caption{Power counting Rules for the $\text{SECT}_{\text{I}}$ and $\text{SECT}_{\text{II}}$ fields \cite{Bauer:2000yr,Bauer:2001yt}.}
\begin{center}
{\begin{tabular}{c|c|c|c}
\hline
\hline Fields &Field Scaling&Fields &Field Scaling\\
\hline
$\xi^{\text{I}}_{n(\bar{n})}$&$\lambda$&$\xi^{\text{II}}_{n(\bar{n})}$&$\eta$\\
$\xi^{\text{I}}_{us}$&$\lambda^3$&$\xi^{\text{II}}_{s}$&$\eta^{3/2}$\\
$(A^{\text{I}}_{n}\cdot n,A^{\text{I}}_{n}\cdot \bar{n},A^{\text{I}}_{n\bot})$&$(\lambda^2,1,\lambda)$&$(A^{\text{II}}_{n}\cdot n,A^{\text{II}}_{n}\cdot \bar{n},A^{\text{II}}_{n\bot})$&$(\eta^2,1,\eta)$\\
$(A^{\text{I}}_{\bar{n}}\cdot n,A^{\text{I}}_{\bar{n}}\cdot \bar{n},A^{\text{I}}_{\bar{n}\bot})$&$(1,\lambda^2,\lambda)$&$(A^{\text{II}}_{\bar{n}}\cdot n,A^{\text{II}}_{\bar{n}}\cdot \bar{n},A^{\text{II}}_{\bar{n}\bot})$&$(1,\eta^2,\eta)$\\
$A_{us}^{\text{I}}$&$\lambda^2$&$A_{s}^{\text{II}}$&$\eta$\\
\hline
\hline
\end{tabular} }
\end{center}\label{powercountingSCET}
\end{table}

Within $\text{NRQCD}$~\cite{Bodwin:1994jh,Luke:1999kz}, there are four typical fields~\cite{Beneke:1997zp}: 1) the Pauli spinor quark field $\psi(\chi)$ with momentum
$p^{\text{NR}}_{\varphi(\chi)}=(E,\vec{p})\sim (\frac{|\vec{q}_{_{B_c}}|}{M_{B_c}}, \vec{q}_{_{B_c}})$; 2) the potential gluon  field $A^{\text{NR}}_{p}$ with momentum
$p^{\text{NR}}_{p}\sim (\frac{|\vec{q}_{_{B_c}}|}{M_{B_c}}, \vec{q}_{_{B_c}})$; 3)
 the soft gluon field $A^{\text{NR}}_{s}$ with  momentum $p_{s}^{\text{NR}}\sim (|\vec{q}_{_{B_c}}|,\vec{q}_{_{B_c}})$; 4) the ultra-soft gluon
 field $A^{\text{NR}}_{us}$ with momentum $p^{\text{NR}}_{us}\sim (\frac{|\vec{q}_{_{B_c}}|}{M_{B_c}},\frac{\vec{q}_{_{B_c}}}{M_{B_c}})$. $\vec{q}_{_{B_c}}$ is the relative momentum between the quark and the anti-quark of the $B_c$ meson. According the recent analysis \cite{Wang:2015bka}, we take $\vec{q}_{_{B_c}}^2\sim 1~\text{GeV}^2 $. Therefore, numerically, we have $\sqrt{(p_{s}^{\text{NR}})^2}\sim\sqrt{ (p_{s}^{\text{NR}})^2}\sim\sqrt{m_b\Lambda_{H}}$.

As to the transition currents $J^{\text{I}}$s, they fall into two categories: 1)
the weak flavor transition currents $J_{w}^{\text{I}}$s, which are induced by
$H_W$; 2) the QCD currents $J_{c}^{\text{I}}$s, which  are
caused by the pure QCD interactions and obtained by integrating out the hard ($\sim m_b$) QCD interactions. According to the number of $J_{c}^{\text{I}}$s, it is convenient to classify the transition amplitudes into two types, $A_{w}$s which are induced by no $J_{c}^{\text{I}}$s,  and $A_{c}$s those are mediated by at least one $J_{c}^{\text{I}}$s.

For describing the DOFs $\sim\Lambda_{H}$, the intermediate fluctuations   $\sim\sqrt{m_b\Lambda_{H}}$ are integrated out. Then, the transition currents $J^{\text{II}}$ and  final effective theory $\text{pNRQCD}+\text{SECT}_{\text{II}}$ are matched onto, corresponding to the $\Lambda_H$ momentum modes.
In the framework of $\text{pNRQCD}$~\cite{Luke:1999kz}, the momentum modes $p_{s}^{\text{NR}}$ and $p_{s}^{\text{NR}}$ are integrated out, leaving only the ultra-soft gluon $A^{\text{NR}}_{us}$ and the Pauli spinor quark field $\psi(\chi)$.
In $\text{SECT}_{\text{II}}$~\cite{Bauer:2003mga,Bauer:2002aj}, similar to the case of $\text{SECT}_{\text{I}}$, there are also three typical momentum regions: 1) the $n$-collinear quarks $\xi^{\text{II}}_n$ and gluons $A^{\text{II}}_n$ with  $p_c\sim m_b(\eta^2,1,\eta)$; 2) the $\bar{n}$-collinear quarks $\xi^{\text{II}}_{\bar{n}}$ and gluons $A^{\text{II}}_{\bar{n}}$ with $p_{\bar{c}}\sim m_b(1,\eta^2,\eta)$; 3)  the soft quarks $\xi^{\text{II}}_{s}$ and gluons $A^{\text{II}}_{s}$ with $p_{s}\sim m_b(\eta,\eta,\eta)$.
Here $\eta=\lambda^2=\Lambda_{H}/m_b$ is the expansion parameter. The field scalings  for these $\text{SECT}_{\text{II}}$ fields \cite{Bauer:2000yr} are also listed in Table.~\ref{powercountingSCET}.

\subsection{The Leading contributions of $A_{w}$}

In this part, we pick out the leading contributions of $A_{w}$. At the scale $\sim \sqrt{m_b\Lambda_H}$, $A_{w}$ is induced by $J_{w}^{\text{I}}$s and the $\text{SCET}_{\text{I}}$  Lagrangian $\mathcal{L}_c$ and $\mathcal{L}_{us}$. Here we have $\mathcal{L}_c=\mathcal{L}^0_{\xi\xi}+\mathcal{L}^1_{\xi\xi}+\mathcal{L}^2_{\xi\xi}+\mathcal{L}^{1}_{\xi q}+\mathcal{L}^{2a}_{\xi q}+\mathcal{L}^{2b}_{\xi q}+\mathcal{L}^0_{cg}+\mathcal{L}^1_{cg}+\mathcal{L}^2_{cg}$. The explicit forms of these $\text{SCET}_{\text{I}}$  Lagrangian can be found in Ref.~\cite{Bauer:2003mga}.
The relevant  $J_{w}^{\text{I}}$s in this work are
\begin{equation}\label{jws}
\begin{split}
J_w^0&= \int \text{d}\omega_2\text{d}\omega_4~\left[C_{w}^{01}(\omega_2,\omega_4)~\left(\chi_{\bar{c}}^{\dag} \Gamma^{01}_{A} \psi_b \right)\left(\bar{q}'_{\bar{n},\omega_2}\Gamma_{B}^{01}q_{n,\omega_4}\right)+C_{w}^{02}\left(\chi_{\bar{c},\beta}^{\dag} \Gamma^{02}_{A} \psi_{b,\alpha} \right)\left(\bar{q}'_{\bar{n},\omega_2,\alpha}\Gamma_{B}^{02}q_{n,\omega_4,\beta}\right)\right]
,\\
J_w^{1}&=\int \text{d}\omega\text{d}\omega_2\text{d}\omega_3~C_{w}^{1}(\omega,\omega_2,\omega_3)~\left(\chi_{\bar{c}}^{\dag} \Gamma^{1}_{\mu}B_{n,\omega}^{\bot\mu} \psi_b \right)\left(\bar{q}'_{\bar{n},\omega_2}\Gamma_{\bar{n}} q_{\bar{n},\omega_3}\right)
,\\
J_w^{2A}&=\int \text{d}\omega_1\text{d}\omega_2\text{d}\omega_3\text{d}\omega_4 ~C_{w}^{2A}(\omega_1,\omega_2,\omega_3,\omega_4)~\left(\chi_{\bar{c}}^{\dag} \Gamma^{2A}_{\mu\nu} \psi_b \right)\left(\bar{q}'_{\bar{n},\omega_2}\Gamma_{\bar{n}} q_{\bar{n},\omega_3}\right)\text{Tr}\left[B_{n,\omega_1}^{\bot\mu}B_{n,\omega_4}^{\bot\nu}\right]
,\\
J_w^{2B}&=\int \text{d}\omega_1\text{d}\omega_2\text{d}\omega_3\text{d}\omega_4~C_{w}^{2B}(\omega_1,\omega_2,\omega_3,\omega_4)~\left(\chi_{\bar{c}}^{\dag} \Gamma^{2B} \psi_b \right)\left(\bar{q}'_{\bar{n},\omega_2}\Gamma_{\bar{n}} q_{\bar{n},\omega_3}\right)
\left(\bar{q}'_{n,\omega_1}\Gamma_{n} q_{n,\omega_4}\right),
\end{split}
\end{equation}
where $\chi_{\bar{c}}$ and $\psi_b$ are the Pauli spinor fields corresponding to the $\bar{c}$ and $b$ quarks, respectively.
$q_{n,\omega_i}$s are  defined as
$q_{n,\omega_i}\equiv[\delta(\bar{n}\cdot\mathcal{P}-\omega_i)W^{\dag}_n\xi_{n}^{\text{I}}]$~\cite{Bauer:2001cu}. $\mathcal{P}$ is the operator picking out the large label momenta. $W_{n}$ is the conventional Wilson line $W_n[{\bar{n}\cdot A^{\text{I}}_{n}}]$ after extracting the phase exponent $e^{-i\mathcal{P}\cdot x}$. $\xi_{n}^{\text{I}}$ is the $n$-collinear field in $\text{SCET}_{\text{I}}$, as introduced in Sec.~\ref{Sec:powercounting}.

In Eq.~\eqref{jws}, $B_{n,\omega}^{\bot}$ is also introduced, which is defined as $B_{n,\omega}^{\bot}\equiv[B_{n}^{\bot}\delta(\bar{n}\cdot\mathcal{P}^{\dag}-\omega)]$. Here we have \cite{Pirjol:2002km}
\begin{equation}\label{definionB}
B_{n}^{\bot\mu}=\frac{1}{g}\left[\frac{1}{\bar{n}\cdot\mathcal{P}}W_{n}^{\dag}[i\bar{n}\cdot D_n,iD_{n}^{\bot\mu}]W_{n}\right],
\end{equation}
where $i\bar{n}\cdot D_n=\bar{n}\cdot\mathcal{P}+g \bar{n}\cdot A_{n}^{\text{I}}$ and $iD_n^{\bot}=\mathcal{P}^{\bot}+g  A_{n}^{\text{I}\bot}$.
Using the building operators $q_{n,\omega_i}$ and $B_{n,\omega}^{\bot}$ to construct the currents is quite convenient, because these building blocks are invariant under the collinear-gauge transformations \cite{Bauer:2001yt}.

Within $\text{SCET}_{\text{I}}$, the scaling of $A_w$ can be expressed as $\lambda^{N_{J}+N_{\mathcal{L}}}(N_{J},N_{\mathcal{L}}\geq0)$. $\lambda^{N_J}$ is the power counting for $J_{w}^{\text{I}}$s. For instance, $\lambda^1$ corresponds to $J_w^{1}$. $\lambda^{N_{\mathcal{L}}}$ stands for the scaling caused by $\text{SCET}_{\text{I}}$  Lagrangian. As a  example, if we consider $A_w$ is induced by the time-product $T\left[ J_w^0,  \mathcal{L}^2_{\xi\xi},\mathcal{L}^{1}_{\xi q} \right]$, then we have $N_{\mathcal{L}}=3$.

If we integrating out the DOFs $\sim\sqrt{m_b\Lambda_H}$, then the $\text{SCET}_{\text{II}}$ are matched onto. According to Ref.~\cite{Bauer:2003mga}, within
$\text{SCET}_{\text{II}}$, the power counting for $A_w$ is $\eta^{(N_{J}+N_{\mathcal{L}})/2+N_{uc}}(N_{uc}\geq0)$. $N_{uc}$ is caused by lowering the off-shellness  of the un-contracted collinear fields.

In this way,  the leading contributions of $A_w$ in $\eta$ can be picked out.
\begin{enumerate}
  \item Case of $N_{J}=0,N_{\mathcal{L}}=0$. Here we show that this kind of amplitudes do not  contribute to the $B_c\to M_1M_2$ processes. As to $B_c\to M_1M_2$ decays, the final mesons involve even $n(\bar{n})$-collinear quarks. However,  as shown in Eq.~\eqref{jws},  there is odd $n(\bar{n})$-collinear quark field. No matter how many $\mathcal{L}^0_{cg}$ and $\mathcal{L}^0_{\xi\xi}$s are contracted with $J_w^0$ , there are still odd final $n(\bar{n})$-collinear quark fields. Therefore, the $B_c\to M_1M_2$ processes do not include this kind of amplitudes.
  \item Case of $N_{J}=1,N_{\mathcal{L}}=0$. Although there are even $n(\bar{n})$-collinear quarks in $J_w^1$, $B_c\to M_1M_2$ transition still receives no contributions from this case. This is because the $n(\bar{n})$ DOF in $J_w^1$ is color-octet. In the leading $\text{SCET}_{\text{I}}$ Lagrangian, namely, $N_{\mathcal{L}}=0$, the $n$ collinear DOFs decouple from the $\bar{n}$ and ultra-soft ones. Thus, the final $n(\bar{n})$ fields are all generated originally from  $B_{n,\omega}^{\bot\mu}$ in $J_w^1$, which makes the final $n(\bar{n})$ meson color-octet. So the $B_c\to M_1M_2$ decays do not contain the amplitudes for this case.
  \item Case of $N_{J}=0,N_{\mathcal{L}}=1$. This case is similar to the $N_{J}=0,N_{\mathcal{L}}=0$ one, which also produces odd $n(\bar{n})$-collinear quarks. Thus, there is no overlapping amplitude for the $B_c\to M_1M_2$  transitions.
  \item Case of $N_{J}=2,N_{\mathcal{L}}=0$. $J_w^{2B}$ will contribute to $B_c\to M_1M_2$ decays. $J_w^{2A}$ contributes only for the isosinglet final states, such as $\eta$, $\eta'(958)$ mesons. Their typical diagrams are plotted in Figs.~1 (a,b).
  \item Case of $N_{J}=0,N_{\mathcal{L}}=2$. In order to produce even $n(\bar{n})$-collinear quarks, only $T[J_w^0, \mathcal{L}^{1}_{\xi_{n} q},\mathcal{L}^{1}_{\xi_{\bar{n}} q}]$ is possible. But in  this time-product, the number of $iD_{n(\bar{n})}^{\bot}$ is odd, which introduces extra suppressions from $N_{uc}$. In this case, $N_{uc}\geq1$. Therefore, at the leading order in $\eta $, the amplitudes for $N_{J}=0,N_{\mathcal{L}}=2$ do not contribute.
  \item Case of $N_{J}=1,N_{\mathcal{L}}=1$. In this case, the product $T[J_w^1,\mathcal{L}^{1}_{\xi q}]$ will not contribute, since it does not produce the even $n(\bar{n})$-collinear quarks. But the products $T[J_w^1,\mathcal{L}^{1}_{\xi  \xi}]$  and  $T[J_w^1,\mathcal{L}^{1}_{cg}]$ do. The examples of these two products are illustrated in Figs.~1 (c,d).
\end{enumerate}

\begin{figure}[htbp]\label{Fig:ABCDAw}
\centering
\subfigure[~Typical diagram for $A_{wA}^2$]{\includegraphics[width =
0.31\textwidth,height=0.15\textheight]{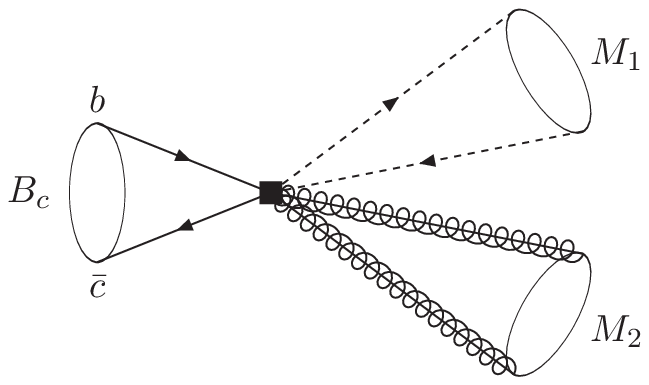}}
\subfigure[~Typical diagram for $A_{wB}^2$]{\includegraphics[width =
0.31\textwidth,height=0.15\textheight]{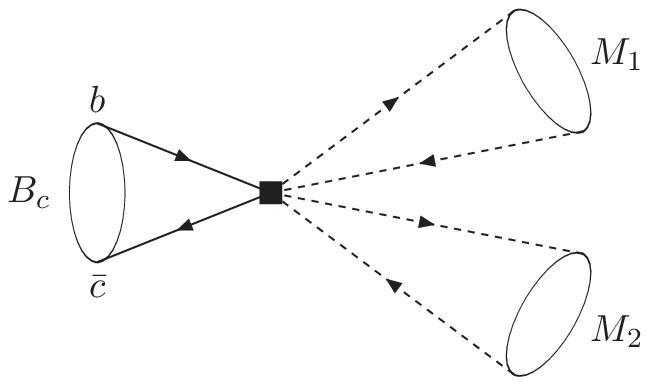}}
\subfigure[~Typical diagram for $A_{wC}^2$]{\includegraphics[width =
0.31\textwidth,height=0.15\textheight]{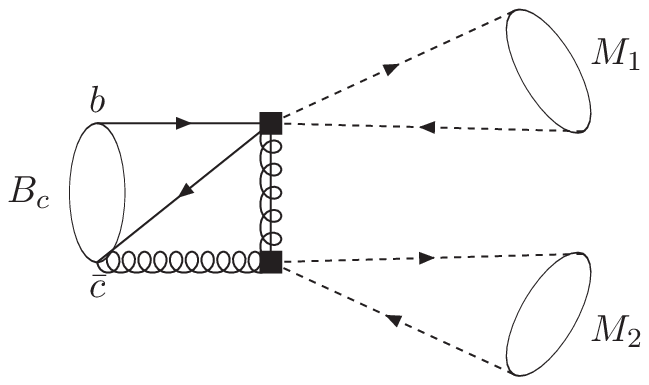}}
\subfigure[~Typical diagram for $A_{wD}^2$ ]{\includegraphics[width =
0.31\textwidth,height=0.15\textheight]{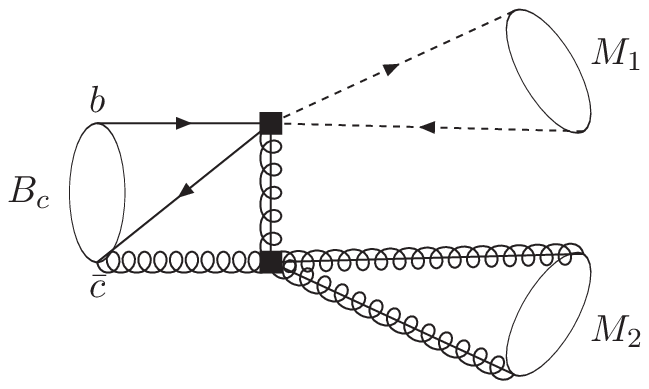}}\caption{Typical diagrams for $A_{wA}^2$, $A_{wB}^2$, $A_{wC}^2$ and $A_{wD}^2$. The solid lines stand for the initial $b(\bar{c})$ quarks, while the dash lines denote the final collinear quarks.  A spring is the (ultra-)soft gluon, but the spring with a line though it represents the collinear gluon. Figs.~(a,d) contribute only to the isosinglet final meson, such $\eta$, $\eta'(958)$.}
\end{figure}

In summary, the operators $J_w^{2A}$, $J_w^{2B}$, $T[J_w^1,\mathcal{L}^{1}_{\xi q}]$  and  $T[J_w^1,\mathcal{L}^{1}_{cg}]$ contribute to the $B_c\to M_1M_2$ processes in the leading order in $\eta$. The according transition amplitudes are
\begin{equation}\label{ampinleadingorder}
\begin{split}
A^2_{wA}&=\langle M_1M_2|J_w^{2A}(0) |B_c^-\rangle,\\
A^2_{wB}&=\langle M_1M_2|J_w^{2B}(0) |B_c^-\rangle,\\
A^2_{wC}&=\langle M_1M_2|\int \text{d}x~ T[J_w^1(0),\mathcal{L}^{1}_{\xi \xi}(x)] |B_c^-\rangle,\\
A^2_{wD}&=\langle M_1M_2|\int \text{d}x~T[J_w^1(0),\mathcal{L}^{1}_{cg}(x) |B_c^-\rangle.
\end{split}
\end{equation}
Consider that in the leading $\text{SCET}_{\text{I}}$ Lagrangian the collinear fields decouple from the ultra-soft fields. Therefore, we have
\begin{equation}\label{factorizationformula1}
    \begin{split}
        A^2_{wA}=&  \int \text{d}\omega_1\text{d}\omega_2\text{d}\omega_3\text{d}\omega_4 ~C_{w}^{2A} (\omega_1,\omega_2,\omega_3,\omega_4)~\langle 0|  \chi_{\bar{c}}^{\dag} \Gamma^{2A}_{\mu\nu} \psi_b   |B_c^-\rangle\langle M_1|  \bar{q}'_{\bar{n},\omega_2}\Gamma_{\bar{n}} q_{\bar{n},\omega_3}  |0\rangle\\
         &\langle M_2|\text{Tr} [B_{n,\omega_1}^{\bot\mu}B_{n,\omega_4}^{\bot\nu} ]|0\rangle,\\
         A^2_{wB}=& \int \text{d}\omega_1\text{d}\omega_2\text{d}\omega_3\text{d}\omega_4 ~C_{w}^{2B}( \omega_1,\omega_2,\omega_3,\omega_4) ~\langle 0|  \chi_{\bar{c}}^{\dag} \Gamma^{2B} \psi_b  |B_c^-\rangle     \langle M_1|  \bar{q}'_{\bar{n},\omega_2}\Gamma_{\bar{n}} q_{\bar{n},\omega_3}  |0\rangle\\
&\langle M_2 |  \bar{q}'_{n,\omega_1}\Gamma_{n} q_{n,\omega_4}  |0\rangle.\\
    \end{split}
\end{equation}

However, there are interactions between the collinear  and ultra-soft fields in
$\mathcal{L}^{1}_{\xi \xi}$ and $\mathcal{L}^{1}_{cg}$. Thus, we have
\begin{equation}\label{factorizationformula2}
    \begin{split}
        A^2_{wC}=&\int \text{d}x\text{d}\omega\text{d}\omega_2\text{d}\omega_3~C_{w}^{1}(\omega,\omega_2,\omega_3)~\langle M_2|T\left[\left(\chi_{\bar{c}}^{\dag} \Gamma^{1}_{\mu} \psi_b B_{n,\omega}^{\bot\mu}\right)(0), \mathcal{L}^{1}_{\xi \xi}(x)\right] |B_c^-\rangle   \langle M_1|  \left(\bar{q}'_{\bar{n},\omega_2}\Gamma_{\bar{n}} q_{\bar{n},\omega_3}\right)|0\rangle,\\
         A^2_{wD}=&\int \text{d}x\text{d}\omega\text{d}\omega_2\text{d}\omega_3~C_{w}^{1}(\omega,\omega_2,\omega_3)~\langle M_2|T\left[\left(\chi_{\bar{c}}^{\dag} \Gamma^{1}_{\mu} \psi_b B_{n,\omega}^{\bot\mu}\right)(0), \mathcal{L}^{1}_{cg}(x)\right] |B_c^-\rangle   \langle M_1|  \left(\bar{q}'_{\bar{n},\omega_2}\Gamma_{\bar{n}} q_{\bar{n},\omega_3}\right)|0\rangle.
    \end{split}
\end{equation}

The typical diagrams of $A^2_{wA}$, $A^2_{wB}$, $A^2_{wC}$ and $A^2_{wD}$ are illustrated in Fig.~1.

\subsection{The analysis of $A^2_{wB}$}

In the last subsection, we prove the factorization formulae of $A^2_{wA}$, $A^2_{wB}$, $A^2_{wC}$ and $A^2_{wD}$.
Here we lay stress on the calculations of $A^2_{wB}$. The analysis of $A^2_{wA}$ can be performed in a similar manner. The estimations of $A^2_{wC}$ and $A^2_{wD}$ involve the non-factorizable  matrix elements $\langle M_2|T\left[\left(\chi_{\bar{c}}^{\dag} \Gamma^{1}_{\mu} \psi_b B_{n,\omega''}^{\bot\mu}\right)(0), \mathcal{L}^{1}_{\xi \xi}(x)\right] |B_c^-\rangle$ and $\langle M_2|T\left[\left(\chi_{\bar{c}}^{\dag} \Gamma^{1}_{\mu} \psi_b B_{n,\omega''}^{\bot\mu}\right)(0), \mathcal{L}^{1}_{cg}(x)\right] |B_c^-\rangle$. We expect them to be determined from the future experimental data
or the non-perturbative method.

For the amplitude $A^2_{wB}$, as shown in Eq.~\eqref{factorizationformula1}, the hadronic matrix elements $ \langle 0|  \chi_{\bar{c}}^{\dag} \Gamma^{2B} \psi_b  |B_c^-\rangle  $, $   \langle M_1|  \bar{q}'_{\bar{n},\omega_2}\Gamma_{\bar{n}}
q_{\bar{n},\omega_3}  |0\rangle$ and $
\langle M_2 |  \bar{q}'_{n,\omega_1}\Gamma_{n}q_{n,\omega_4}  |0\rangle$ are involved.

Considering that the $B_c$ meson is dominated by the $^1S_0^{[1]}$ Fock state, the matrix $\Gamma^{2B}$ should be $I$ and the initial hadronic matrix element can be parameterized as
\begin{equation}\label{BcDC}
   \langle 0|  \chi_{\bar{c}}^{\dag}  \psi_b  |B_c^-\rangle =if_{B_c}M_{B_c},
\end{equation}
where  $f_{B_c}$ is decay constant of the $B_c$ meson.

 As to the final hadronic matrix elements, the matrices $\Gamma_{n}$ and $\Gamma_{\bar{n}}$ are involved. In general,
 they can be represented by the following basis
$$
\{ I, \gamma_5, \not\!n, \not\!\bar{n},\gamma^{\mu}_{\bot},\not\!n\gamma_5, \not\!\bar{n}\gamma_5,\gamma^{\mu}_{\bot}\gamma_5,\not\!n\gamma_{\bot}^{\mu}, \not\!\bar{n}\gamma_{\bot}^{\mu}, (\not\!n\not\!\bar{n}-2) \}.
$$
 If the properties of the $\text{SCET}_{\text{I}}$ fields $\frac{\not\!n ~\not\!\bar{n}}{4}q_{n,\omega_i}=q_{n,\omega_i}$ and $\frac{\not\!\bar{n} ~\not\!n}{4}q_{\bar{n},\omega_i}=q_{\bar{n},\omega_i}$ are considered, only the set of matrices  $\Gamma_{n}=\not\!\bar{n}P_L$ and $\Gamma_{\bar{n}}=\not\!n P_L$ contribute. (The constructions of these local six-quark operators are discussed detailedly in Ref.~\cite{Arnesen:2006vb}. Here we directly use their results.)

 Therefore, we have $   \langle M_1|  \bar{q}'_{\bar{n},\omega_2}\not\!n P_L
q_{\bar{n},\omega_3}  |0\rangle$ and $
\langle M_2 |  \bar{q}'_{n,\omega_1}\not\!\bar{n}P_Lq_{n,\omega_4}  |0\rangle$. According to Ref.~\cite{Bauer:2001cu}, these two hadronic matrix elements are are  just the conventional light cone wave functions in the momentum space.
Based on Ref.~\cite{Beneke:2000wa}, we have
\begin{equation}\begin{split}
\langle P(p)|\bar{q}_{n,\omega_q}\not\!\bar{n}P_Lq'_{n,\omega_{q'}}|0\rangle&=\frac{-if_Pp\cdot\bar{n}}{2}\int_0^1 dx~ \delta(x p\cdot\bar{n}-\omega_q)\delta(\bar{x} p\cdot\bar{n}+\omega_{q'})\phi_{P},\\
\langle V(p)|\bar{q}_{n,\omega_q}\not\!\bar{n}P_Lq'_{n,\omega_{q'}}|0\rangle&=\frac{-if_Vp\cdot\bar{n}}{2}\int_0^1 dx~ \delta(x p\cdot\bar{n}-\omega_q)\delta(\bar{x} p\cdot\bar{n}+\omega_{q'})\phi_{V},
\end{split}\label{PVDC}\end{equation}
where $\bar{x}=1-x$. Usually, $\phi_{P(V)}$ can be expanded in the Gegenbauer polynomials \cite{Beneke:2001ev}
\begin{equation}\label{Gegenbauer}
\phi_{P(V)}=6x(1-x)\left[1+\sum_{n=1}^{\infty}a^{n}_{P(V)}C_{n}^{3/2}(2x-1)\right],
\end{equation}
where $a^{n}_{P(V)}$s are the Gegenbauer moments, which can be obtained from  lattice simulations~\cite{Braun:2015lfa,Arthur:2010xf}. $C_{n}^{3/2}(u)$s are the Gegenbauer polynomials.
In our numerical calculations, we truncate this expansion at $n=2$, using $C_1(u)=3u$ and $C_2=\frac{3}{2}(5u^2-1)$.

Plugging Eqs.~(\ref{BcDC}-\ref{PVDC}) into Eq.~\eqref{factorizationformula1}, $A^2_{wB}$ can be re-written as
\begin{equation}\begin{split}
  A^2_{wB}= \frac{ f_{B_c}f_{M_1}f_{M_2}}{27}\int_0^1 dxdy~C^1_w(x,y)\phi_{M_1}(x)\phi_{M_2}(y).
\end{split}\label{ALocal22}\end{equation}

\begin{figure}[htbp]
\centering{\includegraphics[width =
0.71\textwidth,height=0.15\textheight]{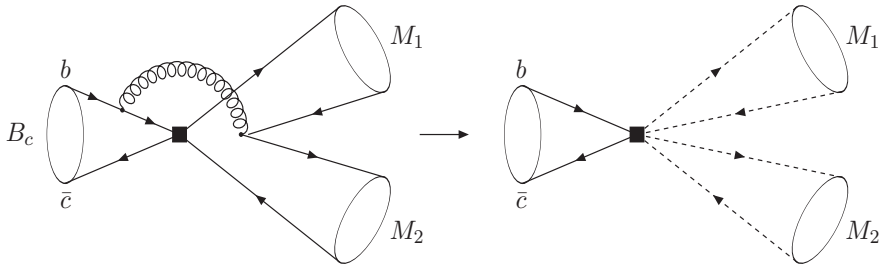}}\caption{Matching procedure for $A^2_{wB}$ in QCD (left diagram) and SCET (right diagram).}
\end{figure}

Matching at tree level, as  shown in Fig.~2, we have
\begin{equation}\label{finalresult}
    C^1_{w}(x,y)=\frac{4\pi G_F\alpha_s(m_b) C_2V_{cb}V^*_{uq}}{\sqrt{2}y(\bar{x}y-\alpha_1\bar{x}-\alpha_1 y+i\epsilon)},
\end{equation}
where $\alpha_{1}=m_{b}/M_{B_c}$. This result is in agreement with the one in Ref.~\cite{DescotesGenon:2009ja}. If we take $\alpha_1\to1$, Eq.~\eqref{finalresult} also agree with the results in Refs.~\cite{Arnesen:2006vb,Beneke:2001ev}.

\subsection{The Leading contributions of  $A_{c}$}\label{Sec:AFH2}
 In this part, we turn to analyzing the leading $A_{c}$ in $\eta$.
$A_{c}$s are induced by one $J_w$ and at least one  $J_c$. From the $\text{SCET}$ power counting rules, at the leading order in $\eta$, $A_{c}^0$ is mediated by $T\left[J_w^0, J_c^0\right]$. The expression of $J_w^0$ has been given in Eq.~\eqref{jws}.
\begin{figure}[htbp]
\centering{\includegraphics[width =
0.71\textwidth,height=0.15\textheight]{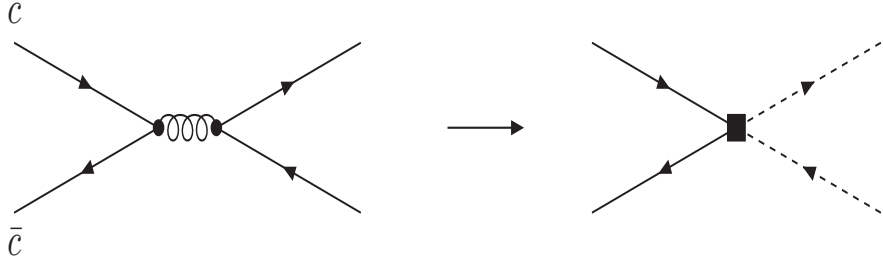}}\caption{Matching procedure for $ J^{0}_{c}$ in QCD (left diagram) and SCET (right diagram).}
\end{figure}
For $J^{0}_{c}$, at the tree level matching, as shown in Fig.~3, we have
\begin{equation}\label{Jc0definition}
\begin{split}
    J^{0}_{c}&=\int\text{d}\omega_1\text{d}\omega_3\left[D_{1}(\chi_{\bar{c}}^{\dag}\sigma_{\bot}^{\mu}\psi_c)(\bar{q}_{n,\omega_1}\gamma_{\bot\mu}q_{\bar{n},\omega_3})
    +D_2(\chi_{\bar{c}\beta}^{\dag}\sigma_{\bot}^{\mu}\psi_{c,\alpha})(\bar{q}_{n,\omega_1\alpha}\gamma_{\bot\mu}q_{\bar{n},\omega_3\beta})\right].
\end{split}\end{equation}
Here $D_1=\frac{2\pi\alpha_s(m_b)}{3\omega_1\omega_3}$ and $D_2=\frac{-2\pi\alpha_s(m_b)}{\omega_1\omega_3}$.

The factorization properties of $\text{SCET}$ yield that $A_{c}^0$ can be re-written as
\begin{equation}\label{Jcfactorization}
\begin{split}
A_{c}^0\propto&\sum_{i,j}\int \text{d}z\text{d}\omega_1\text{d}\omega_2\text{d}\omega_3\text{d}\omega_4~C_{w}^{0i}D_{j}~e^{-i(\omega_1-\omega_3)z}\langle0|T\left\{\left[\chi_{\bar{c}}^{\dag} \Gamma^{01}_{A} \psi_b \right](0),\left[\chi_{\bar{c}}^{\dag}\sigma_{\bot}^{\mu}\psi_c
    \right](z)
    \right\} |B_c^-\rangle  \\
    &  \langle M_1|  \bar{q}'_{\bar{n},\omega_2}\Gamma_{\bar{n}}
q_{\bar{n},\omega_3}  |0\rangle
\langle M_2 |  \bar{q}'_{n,\omega_1}\Gamma_{n}q_{n,\omega_4}  |0\rangle.
\end{split}
\end{equation}
In Eq.~\eqref{Jcfactorization}, the color indices are implicit for readability. The example of this amplitude is illustrated in Fig.~4.

\begin{figure}[htbp]
\centering{\includegraphics[width =
0.31\textwidth,height=0.15\textheight]{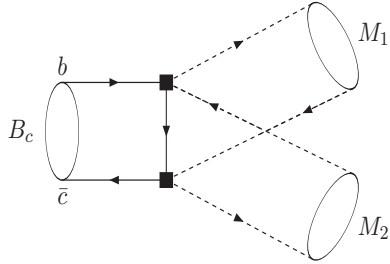}}\caption{Typical diagrams for $A_{c}^0$.}
\end{figure}

Here we interpret the first hadronic term in Eq.~\eqref{Jcfactorization} as the non-perturbative soft functions. This is because that
the soft gluons may be exchanged between the produced $c$
quark and the initial constituent $b(\bar{c})$ quark.

In order to see this, we  approximatively consider $\omega_2=-\omega_3=\omega_1=-\omega_4=M_{B_c}/2$.  In this way, the $c$ quark produced by $J^{0}_{w}$ moves non-relativistically and is almost on-shell. When this $c$ quark and the initial constituent $\bar{c}$ quark are annihilated by $J^{0}_{c}$, it is observed that $(\tilde{P}_c+P_{\bar{c}})^2 \sim M_{J/\psi}^2$. ($\tilde{P}_c$ denotes the momentum of the propagated $c$ quark, while $P_{\bar{c}}$ stands for the initial constituent $\bar{c}$ quark.) Therefore, it is reasonable to expect soft gluons exchanged between the propagated $c$ and the initial partons.

Actually, this situation is not unique in the analysis of $\text{SCET}$. In the $B\to M_1M_2$ processes, there are long-distance charming penguins \cite{Bauer:2004tj}, in which soft gluons are also exchanged among the produced $c$ quarks, the spectator quark and the initial $b$ quark.

\section{Numerical Results and the Discussions}\label{phno}
In this part, we present the numerical results and phenomenal analysis. In Sec.~\ref{inputs}, the inputs  in the  calculations are introduced. Within Sec.~\ref{discussons}, the
 numerical results are shown.

\subsection{Inputs in calculations}\label{inputs}
The masses and lifetimes of the involved mesons are presented in Table. 1. The mass for $b$ quark is taken as $m_b=4.8~\text{GeV}$~\cite{pdg}, while the mass of $c$ quark is used as $m_c=1.6~\text{GeV}$~\cite{pdg}. 
\begin{table}[!htbp]
\caption{Masses and lifetimes.}
\begin{center}
{\begin{tabular}{|c||c|c|c|c|c|c|c|c|c|c|c|c|c|c|c|c|c}
\hline Meson &$B_c$&$\pi$&$K$&$\rho$&$K^*$\\
\hline Mass~\cite{pdg}&$6.3~\text{GeV}$&$0.14~\text{GeV}$&$0.49~\text{GeV}$&$0.77~\text{GeV}$&$0.89~\text{GeV}$\\
\hline Lifetime~\cite{pdg}&$0.51\times10^{-12}\text{s}$&&&&\\
\hline
\end{tabular} }
\end{center}\label{Table:Br}
\end{table}

In Eq.~\eqref{weakeffectiveHamiltonian} and Eq.~\eqref{finalresult}, $\alpha_s$ and the Wilson coefficients $C_1$ and $C_2$ are involved. Here we take $\alpha_s(m_b)=0.22$, $C_1=1.078$ and $  C_2=-0.184$~\cite{Buchalla:1995vs}.

In Eqs.~(\ref{BcDC}-\ref{PVDC}), the decay constants $f_{B_c,P,V}$ and the Gegenbauer moments $a_{1,2}$ are involved. According to Ref.~\cite{Cvetic:2004qg}, we employ $f_{B_c}=0.322~\text{GeV}$. The other inputs are summerized in Table. 2.
\begin{table}[!htbp]
\caption{Decay constants  and the Gegenbauer moments for the light mesons.}
\begin{center}
{\begin{tabular}{|c|c|c||c|c|c|c|c|c|c|c|c|c|c|c|c|c|c}
\hline Meson &$\pi$&$K$&Meson&$\rho$&$K^*$\\
\hline $f_{M}$~\cite{pdg}&$0.130~\text{GeV}$&$0.156~\text{GeV}$&$f_{M}$~\cite{pdg}&$0.208~\text{GeV}$&$0.217~\text{GeV}$\\
\hline $a_1$~\cite{Braun:2015lfa,Arthur:2010xf}&-&$0.0583$&$a_1^{||}$~\cite{Arthur:2010xf}&-&$0.0716$\\
\hline $a_2$~\cite{Braun:2015lfa,Arthur:2010xf}&$0.136$&$0.175$&$a_2^{||}$~\cite{Arthur:2010xf}&$0.204$&$0.145$\\
\hline
\end{tabular} }
\end{center}\label{Table:Br}
\end{table}

\subsection{Numerical results}\label{discussons}

Here we only show the numerical results of $A_{wB}^2$. $A_{wA}^2$ does not contribute to the open flavor final states, while the evaluations of $A_c^0$,  $A_{wC}^2$  and $A_{wD}^2$ involve the non-perturbative hadronic matrix elements.  We leave the  calculations on $A_c^0$,  $A_{wC}^2$  and $A_{wD}^2$ to the future work.

\begin{table}[ph]
\caption{Numerical results of $A_{wB}^2$ in $10^{-10}~\text{GeV}$.}
\begin{center}
{\begin{tabular}{|c|c|c|c|c|c|c|c|c|c|c|c|c|c|c|c|c|c}
\hline  &Results                            \\
\hline $A_{wB}^2(B_c^-\to K^-K^0)$&$-4.50  $\\
 $A_{wB}^2(B_c^-\to K^{*-}K^{0})$&$-5.94  $\\
  $A_{wB}^2(B_c^-\to K^{-}K^{*0})$&$-6.27 $\\
 $A_{wB}^2(B_c^-\to \pi^-\bar{K}^0)$&$-0.89  $\\
 $A_{wB}^2(B_c^-\to \pi^-\bar{K}^{*0})$&$-1.24 $\\
\hline
\end{tabular} }
\end{center}\label{Table:Br}
\end{table}

$A_{wB}^2$ can be obtained from Eqs.~(\ref{ALocal22}-\ref{finalresult}). 
The numerical results are listed in Table.~\ref{Table:Br}. First,
from Table.~\ref{Table:Br}, all of the $A_{wB}$ results are real. This also happens in the local annihilation amplitudes of the $B\to M_1M_2$ decays~\cite{Arnesen:2006vb}. Second, one may note that $A_{wB}^2(B_c^-\to K^-K^0)$ are comparable with the ones of the $ B_c^-\to K^{*-}K^{0}$ and $B_c^-\to K^{-}K^{*0}$
processes, but much larger than the $B_c^-\to \pi^-\bar{K}^0$ and $B_c^-\to \pi^-\bar{K}^0$ cases. This is caused by the suppressed CKM matrix, namely, $V_{us}/V_{ud}\sim \lambda=0.22 $~\cite{pdg}. Third, although our expression of $A_{wB}^2$  is formally identical to the one in Ref.~\cite{DescotesGenon:2009ja}, the results in Table.~\ref{Table:Br} are different from them. In Ref.~\cite{DescotesGenon:2009ja}, the integration in Eq.~\eqref{finalresult} is done with expanding the parameter $\alpha_1$ and take the asymptotic wave functions. However, in this work, the calculations are performed  without these approximations. 

\section{Conclusion}

In this paper, we investigate the $B_c\to M_1M_2$
decays with the framework of (p)NRQCD+SCET.
Our analysis shows that the leading amplitudes for    $B_c\to M_1M_2$ processes include $A_{wA}^2$, $A_{wB}^2$, $A_{wC}^2$, $A_{wD}^2$ and $A_c^0$.

As to $A_{wA}^2$ and $A_{wB}^2$, from the SCET properties, they can be factorized into the following form
\begin{equation}\label{conFA}
    H\otimes\Phi_{B_c}\otimes\Phi_{\bar{n}}\otimes\Phi_{n}.
\end{equation}
Here $H$ denotes the hard kernel, while  $ \Phi_{B_c}$ and  $\Phi_{n(\bar{n})}$ stand for the initial and final wave functions, respectively. This factorization formulae is in agreement with the PQCD \cite{Liu:2009qa,Yang:2010ba} and QCDF \cite{DescotesGenon:2009ja} results. And our result on $A_{wB}^2$ is formally identical to Ref.~\cite{DescotesGenon:2009ja}.

But for  $A_{wC}^2$, $A_{wD}^2$ and $A_c^0$, the situations are different.
 The amplitude $A_c^0$ includes the initial soft functions, while the ones $A_{wC}^2$ and  $A_{wD}^2$ involve the lagrangian $\mathcal{L}^{1}_{\xi \xi}$ and $\mathcal{L}^{1}_{cg}$, where  the collinear fields are tangled with ultra-soft gluons. Therefore, we expect the amplitudes $A_{wC}^2$, $A_{wD}^2$ and $A_c^0$ can not be expressed as the form in Eq.~\eqref{conFA}.

%
%

\section*{Acknowledgments}

This work was supported in part by the National Natural Science Foundation of China
(NSFC) under Grant No. 11575048, No. 11405037 and No. 11505039. T.Wang was also
supported by PIRS of HIT No.B201506.

\clearpage

\appendix

\section{Details on the $A_{wB}^2$ calculation}\label{details}

In this part, we introduce the details on the $A_{wB}^2$ calculation. From Eq.~\eqref{ALocal22}, it is observed that the numerator of the integrand  is a  polynomial of $x$ and $y$. Hence, we can expand $A_{wB}^2$ in terms of $I(m,n)$s, namely, $A_{wB}^2=\sum_{m,n=0}^{\infty}\mathcal{B}(m,n)I(m,n)$. $\mathcal{B}(m,n)$ is the according parameter, while the elemental integration $I(m,n)~(m,n\geq0)$ is defined as
\begin{equation}\label{eleIntegration}
I(m,n)=\int_0^1 dxdy~\frac{x^my^n}{xy-\alpha_1 x-\alpha_1 y+i\epsilon}.
\end{equation}
From Eq.~\eqref{eleIntegration}, we see  $I(m,n)=I(n,m)$. Hence, in the following paragraphes only $I(m,n)~(m\geq n\geq0)$ is introduced. The case for $n> m>0$ can be obtained from the symmetries.

For the term $I(0,0)$, we have
\begin{equation}\label{I00}\begin{split}
    I(0,0)=-\text{Li}_2\left(-\frac{(1-\alpha_1 )^2}{-\alpha_1 ^2+i \epsilon }\right)+\text{Li}_2\left(\frac{(1-\alpha _1) \alpha_1 }{-\alpha_1 ^2+i \epsilon }\right)+\text{Li}_2\left(-\frac{(-1+\alpha_1 ) \alpha _1}{-\alpha _1^2+i \epsilon }\right)-\text{Li}_2\left(-\frac{\alpha_1 ^2}{-\alpha_1 ^2+i \epsilon }\right).
\end{split}\end{equation}
It seems that the analysis from the Landau equations \cite{Landau:1959yya,Landau:1959fi} implies the end-point singularities in $I(0,0)$. But a careful study shows that those singularities are not in the principal sheet. Hence, $I(0,0)$ is finite. Compared with other $I(m,n)$s, it is observed that $I(0,0)$ is the most singular term. Thus,  all  $I(m,n)$s are also finite. This conclusion agrees with  Ref.~\cite{DescotesGenon:2009ja}.

For the term $I(m,0)~(m\geq1)$, we have
\begin{equation}\label{Im0}\begin{split}
    I(m,0)=\alpha^nI(0,0)+\int^{1-\alpha_1}_{-\alpha_1}du~\sum_{i=0}^{n-1}\left(C_n^iu^{n-1-i}\alpha_1^i\right)\left[\text{Log}(u-\alpha_1 u-\alpha_1^2+i\epsilon)-\text{Log}(-\alpha_1 u-\alpha_1^2+i\epsilon)\right],
\end{split}\end{equation}
where $C_n^i$ is the  binomial coefficient.

As to the term $I(m,n)~(m\geq n\geq 1)$, it is
\begin{equation}\label{Imn}\begin{split}
    I(m,n)=\sum_{j=0}^{n}C_{n}^j\alpha_1^nI(n-j,m-n+j)+\int^1_0dxdy~\sum_{i=0}^{n-1}C_{n}^{i}y^{m-n}(\alpha_1 x+\alpha_1 y)^i(xy-\alpha_1 x-\alpha_1 y)^{n-1-i}
\end{split}\end{equation}
We can evaluate this equation inductively, because the powers of $I(n-j,m-n+j)$s are no more  than $m$. For instance,
$I(1,1)=2\alpha_1 I(1,0)+1,
$ where $I(1,0)$ can be computed from Eq.~\eqref{Im0}.

Consequently, based on Eqs.~(\ref{I00}-\ref{Imn}), all of $I(m,n)$s can be evaluated. The use of  these $I(m,n)$s are quite general. They can not only be employed to calculate Eq.~\eqref{ALocal22}, if we make proper replacements of $\alpha_1$, they are also useful in the calculations of  Ref.~\cite{DescotesGenon:2009ja}.

\end{document}